\documentclass[11pt]{article}
\usepackage{graphicx}
\usepackage{color}
\usepackage{amssymb}
\usepackage{amsmath}
\usepackage{multirow}
\usepackage{epsfig}
\usepackage{amscd}
\def\Journal#1#2#3#4{{#1} {\bf #2}, (#3) #4}

\def\etal{{\it et al.}}

\def\AIP{\em AIP Conf.Proc.}

\def\APJ{\em ApJ.}

\def\EPL{\em Europhys. Lett.}
\def\FOP{\em Found. Phys.}

\def\IMA{{\em Int. J. Mod. Phys.} A}
\def\IMD{{\em Int. J. Mod. Phys.} D}

\def\JHE{\em J. High Ener. Phys.}

\def\JPC{\em J. Phys.: Conf. Series}

\def\MPL{{\em Mod. Phys. Lett.} A}
\def\MRA{\em MNRAS}

\def\NAM{\em Nature Mater.}

\def\NAT{\em Nature}

\def\NPB{{\em Nucl. Phys.} B}

\def\PLB{{\em Phys. Lett.} B}
\def\PNS{\em Pub. Nation. Acad. Sci.}

\def\PRD{{\em Phys. Rev.} D}
\def\PRL{\em Phys. Rev. Lett.}
\def\PRV{\em Phys. Rev.}

\def\PRS{\em Proc.Roy.Soc.Lond.}

\def\RMP{\em Rep. Mod. Phys.}

\def\be{\begin{equation}}
\def\ee{\end{equation}}
\def\bea{\begin{eqnarray}}
\def\eea{\end{eqnarray}}

\title{\bf Issues with vacuum energy as the origin of dark energy}
\author{{\it Houri Ziaeepour} \\
{\small Max Planck Institut f\"ur Extraterrestrische Physik (MPE),} \\
{\small Giessenbachstra$\mathbf{\beta}$e, Postfach 1312, 85741 Garching, Germany.} \\
{\tt houriziaeepour@gmail.com}}

\begin{document}
\maketitle
\begin {abstract}
In this letter we address some of the issues raised in the literature about the conflict between 
a large vacuum energy density, apriori predicted by quantum field theory, and the observed 
dark energy which must be the energy of vacuum or include it. We present a number of arguments 
against this claim and in favour of a null vacuum energy. They are based on the following 
arguments: A new definition for the vacuum in quantum field theory as a frame-independent 
coherent state; Results from a detailed study of condensation of scalar fields in FLRW 
background performed in a previous work; And our present knowledge about the Standard Model of 
particle physics. One of the predictions of these arguments is the confinement of nonzero 
expectation value of Higgs field to scales roughly comparable with the width of electroweak 
gauge bosons or shorter. If the observation of Higgs by the LHC is confirmed, accumulation of 
relevant events and their energy dependence in near future should allow to measure the spatial 
extend of the Higgs condensate.
\end {abstract}


\section {Introduction} \label{sec:intro}
This work is an attempt to find an answer for the following question:
\begin{center}
{\bf \it Can the observed dark energy be due to a small nonzero vacuum energy ?}
\end{center}
This question is not simple to answer. We need to have a precise definition of what we call 
{\it vacuum}. Remind that in quantum field theory both an empty state i.e. a state with no 
particle and the minimum of the potential are called {\it vacuum}. Because energy is the source 
for gravity, a detailed examination of the nature of vacuum and its energy density apriori 
needs a quantum gravity model and a knowledge of the quantum state of the Universe, at least 
from the earliest times relevant for present observations, presumably since inflation inflation epoch. 
String theory, 
which for the time being is the most popular quantum gravity candidate, allows some 
$\sim 10^{500}$ {\it vacuum} states, generated by the compactification of internal dimensions. 
In effective field theory models they correspond to the minima of modulies potential, thus they 
are not {\it empty states}. For the time being there is no generally accepted selection rule 
or probability distribution for this landscape. Considering a more phenomenological setup, for 
instance an empty Minkowski spacetime at infinite past, quantum fluctuations can create pair 
of particles. This process is irreversible because the presence of particles modifies the 
geometry of the spacetime and the probability of self-annihilation, and leads to a non-empty 
state. Therefore, it seems that when gravitational effects are considered, a true {\it vacuum} - 
empty space - cannot exist and is only an abstract concept.

In a few recent reviews of dark energy models~\cite{descalar,devacuumrev,devacuumrev0} authors 
have argued in favour or against an small but nonzero value of vacuum energy density. The 
present letter can be classified in the latter category i.e. here we want to argue that based 
on fundamental principles of quantum mechanics, vacuum energy must be zero. Notably, we show 
that the Casimir effect which is usually used as an evidence for gravitational interaction of 
vacuum~\cite{weinbergde} is related to close relation between quantum mechanics, symmetries, 
and observables. After renormalization these effects are included in the effective mass and 
couplings of particles and the vacuum itself can be considered as being an empty space without 
any energy.

Another issue that is claimed to be problematic for quantum field theory is the large vacuum 
expectation value of known or expected condensates in the framework of Standard Model. In 
particular, quark-antiquark condensate $\langle \bar{q}q\rangle \neq 0$ responsible for the 
breaking of chiral symmetry in QCD~\cite{qcdchiral,qcdchiral0,qcdchiral1} and Higgs condensate 
which is necessary for 
generation of the mass of electroweak gauge bosons~\cite{weinbergde,devacuumrev,devacuumrev0}
have been the focus of a large number of studies. Apriori one of the conditions for the formation 
of condensates is the extension of condensate field to an infinite volume with roughly the same 
amplitude. Apparently, the application of this rule would mean that the Standard Model condensates 
should contribute to the cosmological vacuum energy term in the Einstein equation and should make it 
some $10^{40}$ or more folding larger than the observed dark energy density. Therefore, the small 
observed value for dark energy is in strong contradiction with this expectation, and apriori 
one could consider this as the signature of the failure of quantum field theory. However, 
recently various arguments have been suggested to prove that in the case of quark-antiquark 
condensate, the nonzero expectation value $\langle \bar{q}q\rangle \neq 0$ is confined into 
hadrons volume, and therefore its effect is included in their mass and does not contribute to 
dark energy~\cite{qcdchiralvacu}. Here we extend these arguments to Higgs and other scalar fields 
expected in the extensions of the Standard Model such as SUSY. We show that they cannot have an 
effect on the vacuum energy of the Universe at large scales. Thus, the origin of dark energy must 
be searched elsewhere, notably in the condensation of one or multiple light or preferentially 
massless quantum fields - to prevent the need for fine-tuning.

\section {Vacuum energy in quantum field theories} \label{sec:vacuum}
We begin with a review of the definition of what is called a {\it vacuum}. As everybody knows, 
the dictionary definition of this word is {\it emptiness}, the absence of any material. 
This is exactly the way it is used in the context of classical physics. However, in quantum field 
theory its meaning is more subtle, and for this reason, in the literature its exact definition is 
somehow context dependent. For instance, consider the case of bosonic fields. The Fock space is 
generated by creation and annihilation operators defined according to the following 
decomposition:\footnote{Through out this work we assume $\hbar = c = 1$.}
\bea
&&\phi (\vec{x},t) = \sum_{\{\alpha\}} u_{\{\alpha\}} (\vec{x},t) a_{\{\alpha\}} + 
u^*_{\{\alpha\}} (\vec{x},t) a_{\{\alpha\}}^\dagger \label{phidecomp} \\
&&[a_{\{\alpha\}},a_{\{\alpha'\}}] = 0, \quad\quad 
[a_{\{\alpha\}}^\dagger,a_{\{\alpha'\}}^\dagger] = 0, \quad\quad 
[a_{\{\alpha\}},a_{\{\alpha'\}}^\dagger] = \delta_{\{\alpha\}\{\alpha'\}}
\label{commutrel}
\eea
where $u_{\{\alpha\}}$ and $u^*_{\{\alpha\}}$ are solutions of dynamic equation for the set of 
parameters ${\{\alpha\}}$ in an arbitrary spacetime. The vacuum state is defined as:
\be
a_\alpha~|0\rangle = 0, \quad \forall \alpha \label {vacdef}
\ee
Through this work we use this definition for the vacuum. The decomposition (\ref{phidecomp}) and 
definition (\ref{vacdef}) are usually performed for free fields, thus in presence of 
interactions this decomposition is valid only if the model is perturbative. For instance, in flat 
Minkowski spacetime $\{\alpha\} = \{k^0,\vec{k}\}$, $u_k = e^{-ikx}$ and ${k^0}^2 = \vec{k}^2 + m^2$ 
for on-shell modes. This definition uniquely determines the vacuum, up to an arbitrary Lorentz 
transformation of the reference frame.

For a free classical scalar field the energy density - the $00$ component of energy-momentum 
tensor - is $T^{00} = 1/2 \partial^0 \phi \partial^0 \phi + 1/2 m^2 \phi^2$. For a quantum scalar 
field the same expression is usually used but the classical field is replaced by the quantum field 
operator. Then, the decomposition (\ref {phidecomp}) is applied to determine the vacuum 
expectation value of energy density $\langle 0|\hat{T}^{00}|0\rangle$. Restricting ourselves to 
on-shell modes we find:
\be
\langle 0|\hat{T}^{00}|0\rangle = \langle 0|\frac{1}{(2\pi)^3} \int d^3k~u_ku^*_k 
\omega_k (a_k a_k^\dagger + a_k^\dagger a_k)|0\rangle = \frac{1}{(2\pi)^3} \int d^3k~
\omega_k \rightarrow \infty \quad\quad \omega_k = \sqrt {\vec{k}^2 + m^2} 
\label{t00minkovski}
\ee
The infinite or very large value of $\langle 0|\hat{T}^{00}|0\rangle$ - when a cutoff to UV 
limit of integral in (\ref{t00minkovski}) is introduced - is the famous problem of vacuum 
energy. In curved spacetimes expressions for $u_{\{\alpha\}}$ and $u^*_{\{\alpha\}}$ are usually more 
complicated and in fact exceptionally analytic expression for them exist. The energy $\omega_{\alpha}$ 
is also in general more complicated and depends on time. In some cases such as Anti-de Sitter spacetime, 
after renormalization a finite vacuum expectation value is obtained~\cite{adst00} which is 
inversely proportional to the volume of the spacetime.

The origin of the singularity in (\ref{t00minkovski}) is well understood~\cite{qftcurve}. It is 
due to the fact that in quantum field theory, operators $\phi^2$ and 
$(\partial^0 \phi)^2 \equiv \Pi^2$ in $\hat{T}^{00}$ are not well defined and must be replaced by 
ordered operators:
\bea
\phi^2(x) & \rightarrow & :\phi^2 (x):~\equiv~\lim_{y \rightarrow x}~\biggl \{ \phi (x) 
\phi (y) - \langle 0|\phi (x) \phi (y)|0\rangle \biggr \} \label{orderop} \\
\Pi^2(x) & \rightarrow & :\Pi^2 (x):~\equiv~\lim_{y \rightarrow x}~\biggl \{ \Pi (x) \Pi (y) - 
\langle 0|\Pi (x) \Pi (y)|0\rangle \biggr \} \label{orderopmom}
\eea
This operation brings the vacuum energy to zero, as expected for an empty space, and is one of the 
regularization methods used in quantum field theories. There are other regularization and 
renormalization methods too, some of them more suitable for applications in curved spacetimes, 
see e.g.~\cite{qftcurve} for a review. Nonetheless, generally it is considered that the application 
of regularization and renormalization techniques to energy-momentum tensor is 
ad hoc~\cite{descalar}. Therefore in a mysterious way, vacuum energy of various components must 
cancel each others to make the energy density of the vacuum consistent with observations, 
see~\cite{weinbergde} for a review. Another solution to this problem associates the small 
observed value of dark energy to one of many possibilities offered by the very large landscape of 
string theories~\cite{stringlandscape,stringlandscape0}. But, giving the fact that no generally 
accepted probability distribution on the string landscape exists, the value of vacuum energy density 
must have an anthropological bias~\cite{antropologic,antropologic0}, i.e. universes with large vacuum 
density cannot develop galaxies, stars, and life.

\subsection{New interpretations} \label{sec:newinterp}
Now let's go back and look through the problem from another angle. Mode functions $u_k$ and $u^*_k$ 
are solutions of the field equation which is a differential equation. They include arbitrary 
integration constants that are determined from initial or boundary conditions. In Minkowski 
spacetime conjugate functions $u_k$ and $u^*_k$ are independent solutions of the field equation, 
thus there is no ambiguity in the definition of particles and anti-particles. This is due to the 
fact that in Minkowski space there is a Killing vector for the whole space, and thereby there is 
a {\it natural} vacuum definition. However, in expanding spaces such as FLRW and de Sitter there 
is not a unique {\it natural} vacuum and $u_k$ and $u^*_k$ define vacua which are only 
approximately similar to Minkowski - adiabatic - vacuum. Various vacua are related to each other 
by a Bogolubov transformation, and correspond to adiabatic vacua in frames moving with respect to 
each other, not necessarily with constant velocity. The vacuum state in one frame is a non-vacuum 
state with infinite number of particles in another frame. According to general relativity there is 
no preferred frame, thus there is no preferred vacuum either, see e.g.~\cite{qftcurve} and 
references therein for more details. In the next section we propose another definition for vacuum 
in the context of quantum field theories. We believe it presents better the properties of what 
is called {\it vacuum} in quantum field theories in which vacuum is very far from being an static 
empty space. Before presenting the new definition, we first discuss the physical interpretation 
of singularity in equation (\ref{t00minkovski}).

Consider the operator $a_k a_k^\dagger + a_k^\dagger a_k$ in (\ref{t00minkovski}). The second term is 
the number operator $\hat {N}_k$ and 
its eigen value is the number of particles in mode $k$. By definition the vacuum state $|0\rangle$ 
is an eigen state of $\hat {N}_k~\forall~k$ with eigen value zero. Thus, its application does not 
modify the vacuum. Assuming that an unambiguous notation of particle and anti-particle exists or 
defined by convention with respect to a given frame, the second term in the expression above can be 
decomposed to creation of one particle in mode $k$ and then its annihilation. In a classical view 
these two operations are opposite to each other and should leave the space unchanged - specially 
when gravity and change of geometry due to particle creation discussed in the Introduction is not 
considered. Therefore, the constant term due to noncommutation of $a_k$ and $a_k^\dagger$ can be 
considered as a {\it remnant} energy. One interpretation is that this remnant is due to an error 
aroused from using the classical expression for $T^{\mu\nu}$ which includes ambiguous operators. In 
this case the operator ordering or other regularization schemes are legitimate, and produce the 
correct expression for $T^{\mu\nu}$ in the context of quantum field theories. 

We can also interpret the remnant from a purely quantum mechanical point of view. The expression 
(\ref{phidecomp}) can be decomposed in real space to creation and annihilation operators at the 
point $x$ rather than in Fourier space. This simply corresponds to separately sum over all 
creation and annihilation terms. Then, the decomposition can be interpreted as the following: 
Because we know exactly the position of the created (annihilated) particle, we can have no 
information about its momentum which can be any value including infinity, and thereby the 
singularity of (\ref{t00minkovski}). In addition, we note that creation of a particle at a 
specified point breaks the translation symmetry - Poincar\'e group - of the spacetime, specially 
if we consider its gravitational effect on the spacetime. In fact even without creation of any 
particle the symmetry of the spacetime breaks by just associating an origin to it. The reason is 
again the Heisenberg uncertainty rules. Singling out one point of the spacetime induces infinite 
uncertainty in energy and momentum measurements, thus singularity of 
(\ref{t00minkovski})\footnote{These arguments are in spirit the same as the arguments given by 
Eppley \& Hannah~\cite{semiqmgrincon} and in~\cite{houriqgr} to prove the inconsistency of a 
classical gravity and a quantic matter.}. In this interpretation regularization of 
the integral in (\ref{t00minkovski}) by imposing an energy cutoff is equivalent to introducing an 
uncertainty on the position measurement. In the literature usually the high energy cutoff is 
associated to the high energy physics. Here we see that according to this interpretation it 
presents the highest energy scale or equivalently smallest distances in which the observer can 
verify the presence of a vacuum. This minimum distance depends on the strength of $\phi$ field 
couplings, stronger the coupling longer the distance. This brings us to the issue of the 
definition of vacuum and how its presence can be tested. We discuss this in the next section. 
Meanwhile, we should emphasis that the regularizing cutoff only provides an upper limit on the 
vacuum energy density and does not mean that the latter is nonzero. In this interpretation the 
effect of high energy modes are already considered in the effective mass of the particle and 
should not be considered. Note also that these interpretations do not take into account the 
arbitrary integration coefficients which are implicitly included in $u_k$ and $u^*_k$. Neither we 
used explicit expressions for these functions which are expected to depend on the geometry of 
spacetime. Therefore, these arguments are valid for any vacuum and any spacetime. 

\subsection{Vacuum as a coherent state} 
According to the definition of vacuum in equation (\ref{vacdef}), there is no particle in a vacuum 
state in the frame for which it has been defined. Therefore, we expect no effect on a particle 
that passes through it. This means that such an experience can be used to test the presence of a 
vacuum. However, we know that quantum fluctuations affect a particle and after renormalization 
its effective mass for an observer at rest in the vacuum frame depends on its energy. Therefore, 
it is the sensitivity of an observer to energy variation that determines how well the vacuum can 
be detected. This interpretation is consistent with our claim in the Introduction that vacuum is 
an abstract concept and in the context of quantum field theory its presence is always an 
approximation. Considering the complexity of the vacuum and the fact that it is not really the 
no-particle state that its definition pretend, here we suggest a new definition for it. 

In~\cite{houricondense} we described a generalized coherent state for a scalar field based on 
an original suggestion by~\cite{condwave} as:
\bea
&& |\Psi_{GC}\rangle \equiv \sum_k A_k e^{C_k a_k^{\dagger}} |0\rangle = \sum_k A_k 
\sum_{i=0}^{N \rightarrow \infty} \frac {C_k^i}{i!}(a_k^{\dagger})^i |0\rangle \label{condwaveg} \\ 
&& a_k |\Psi_{GC}\rangle = C_k |\Psi_{GC}\rangle \quad \quad 
\langle \Psi_{GC}| N_k |\Psi_{GC}\rangle = |A_k C_k|^2 \label{condwavegann}
\eea
For $\{C_k \rightarrow 0~\forall~k\}$ this state is neutralized by all annihilation operators and 
the expectation value of particle number approaches zero for all modes. Therefore, this state 
satisfies the condition (\ref{vacdef}) for a vacuum state. Coefficients $A_k$ are relative 
amplitude of modes $k$ and can be nonzero even for vacuum state. In addition, one can extend this 
definition by considering products of $|\Psi_{GC}\rangle$ states. Such a state includes products of 
states in which particles do not have the same momentum. Thus, it consists of all 
combinations of states with any number of particles and momenta. 
\be
|\Psi_{G}\rangle \equiv \sum_{k_1,k_2,\cdots} \biggl (\prod_{k_i} A_{k_i} \biggr) e^{\sum_i C_{k_i} 
a_{k_i}^{\dagger}} |0\rangle \label{condwave}
\ee
In this case all $C_{k_i} \rightarrow 0~\forall~k_i$ in a vacuum state. Under a Bogolubov 
transformation this state is projected to itself:
\be
a_{k_i} = \sum_j \sum_{k_j} {\mathcal A}_{k_j k_i} a'_{k_j} + \sum_j \sum_{k_j} 
{\mathcal B}_{k_j k_i} {a'}^\dagger_{k_j} \quad \quad a^\dagger_{k_i} = \sum_j \sum_{k_j} 
{\mathcal A}^*_{k_j k_i} {a'}^\dagger_{k_j} + \sum_j \sum_{k_j} {\mathcal B}^*_{k_j k_i} a'_{k_j}
\label {bogoltrans}
\ee
Replacing $a_{k_i}^{\dagger}$ in (\ref{condwave}) with the corresponding expression in 
(\ref{bogoltrans}) leads to an expression for $|\Psi_{G}\rangle$ similar to (\ref{condwave}) but 
with respect to the new operator ${a'}^\dagger_k$ and $C'_{k_j} = \sum_i \sum_{k_i} 
{\mathcal A}^*_{k_j k_i} C_{k_i}$. For $C_{k_i} \rightarrow 0~\forall~k_i$ and finite 
${\mathcal A}^*_{k_j k_i}$, $C'_{k_i} \rightarrow 0~\forall~k_i$. Note that here we assume that the 
Bogolubov transformation changes $|0\rangle$ to a similar state which is neutralized by 
$a'_k~\forall~k$. Therefore, in contrast to the null state of the Fock space, $|\Psi_{G}\rangle$ 
is frame-independent. However, it is easy to verify that this new definition of vacuum does 
not solve the problem of singular integral when one tries to determine the expectation value of 
$\hat{T}^{00}$ without operator ordering, because as explained above this issue is related to the 
definition of $\hat{T}^{00}$. Nonetheless, it gives a better insight into the nature of the problem. 
Notably, one can use the number operator $\sum_k \hat{N}_k$ to determine the energy density of 
vacuum because in contrast to $|0\rangle$, the new vacuum $|\Psi_{G}\rangle$ is frame-independent 
and is neutralized by the number operator $\hat {N}_k |\Psi_{G}\rangle = 0 ~\forall~k$. This 
alternative to $\hat{T}^{00}$ for measuring the vacuum energy density has been discussed 
in~\cite{qftcurve}, but has been considered to be a poor replacement because vacuum state in one 
frame can be a state with nonzero number of particles in another frame. Consideration of 
$|\Psi_{G}\rangle$ as vacuum has the advantage that it is invariant under Bogolubov transformation. 
Thus, if $\sum_k |A_k C_k|^2 \rightarrow 0$ in one frame it is null in all frames. If a state do not 
have any particle for all observers, its energy density must be zero. This frame-independent 
argument proves that the application of regularization when one uses $\hat{T}^{00}$ for the same 
purpose is not an ad hoc operation.

Coefficients $A_k$ should be calculated from the full Lagrangian of the model using usual quantum 
field theory techniques. Evidently, the full solutions of propagators which is necessary for 
determining $A_k$'s depend on the initial or boundary conditions. Nonetheless, as 
$|\Psi_{G}\rangle$ contains all possible states, different initial conditions project it to itself. 
In this sense this state is unique. Moreover, as $|\Psi_{G}\rangle$ is a maximally coherent state 
and its member states have vanishing amplitude, they are not directly observable. Similar to a 
usual entangled state, one can say that vacuum {\it collapses} to one of its member states when 
it is observed i.e. in the process of interaction, for instance with an on-shell particle. But 
because any single state has a vanishing amplitude, one can always consider the vacuum unchanged 
even when one or any finite number of its members interact with an untangled state, and thereby is 
indirectly observed as virtual particles. Their effect manifests itself as scale dependence of mass 
and couplings of the field. Thus, like usual definition of vacuum, interactions 
modify properties of the external (untangled) particles at scales relevant for the interaction, but 
they don't change $|\Psi_{G}\rangle$\footnote{In the interaction picture states are evolved 
according to free Hamiltonian. Therefore, interactions do not modify $|\Psi_{G}\rangle$.}. 

Note that $|\Psi_{G}\rangle$ includes all states at any scale. However, in every experiment only 
a limited range of scales are available to observers. They are limited from IR side by the size of 
the apparatus or observational limits such as a horizon, and from UV side by the available energy 
to the observer. The presence of a particle at a given scale i.e. discrimination between vacuum 
and non-vacuum at that scale depends on the uncertainties of distance/energy measurements. At large 
distance scales the limited sensitivity of detectors cannot detect interaction of untangled 
particles with very low energy virtual particles and no violation of energy-momentum conservation 
occurs. This couldn't be true if the vacuum had a large energy-momentum density which could be 
exchanged with on-shell particles at any distant scale.

Does the coherent vacuum state $|\Psi_{G}\rangle$ gravitate ? A detailed answer to this question 
needs a quantum description for gravity and what a vacuum means in this context. Nonetheless, 
even without knowing the details of the vacuum state $|\Psi_{G}\rangle$ for quantum gravity, by 
definition states that make up this coherent state are not observable except when they are 
decohered/collapsed. And when this happens, they will not appear as vacuum. Consequently, they cannot 
influence observations in any way including gravitationally. In a semi-classical view, one expects 
that the expectation number of particles with a given energy and momentum determines the strength 
of gravitational force. Equation (\ref{condwavegann}) shows that this number for any value of 
energy and momentum is null when $\{C_k \rightarrow 0~\forall~k\}$. Thus this state does not feel 
the gravity. This shows the unphysical nature of energy-momentum tensor singularity when its 
classical definition is used in quantum field theory without regularization. 

A good example of such vacuum is the ensemble of electrons in valance energy levels inside a solid.
Although electrons are fermions and their number in each energy level is restricted to 2, this 
system presents very well the properties of the vacuum state discussed above. At zero temperature all 
the states under Fermi level are filled and pair of electrons make an entangled ensemble. Injected 
external - on shell - electrons or photons can decohere e.g. one electron and make a pair of 
electron-hole - an exciton~\cite{exciton}. But the vacuum stays (approximately) unchanged. If 
the injected energy does not significantly increase the temperature, the lifetime of the 
electron-hole pair (exciton) will be short, and they recombine quickly. This is similar to vacuum 
fluctuation of fundamental quantum fields~\footnote{Note that similar to production of Cooper pairs 
of electrons in superconductors, the pairing of electron-hole and formation of the exciton 
pseudo-scalar field occurs only at low temperatures. At higher temperatures electrons and holes 
behave independently as fermionic gases.}. The energy necessary for the creation of electron-hole 
pair is provided by the exciting agent, a photon or another electron, and the effect can be 
considered as a change in the kinetic energy of the colliding photons - usually from a laser 
source. Therefore, creation of virtual particles do not affect the interaction of the system with 
gravity (also see Sec. \ref{sec:virtual}). This example shows that the state $|\Psi_{G}\rangle$ 
unifies the definition and presentation of vacuum state in particle physics and in condense matter 
physics. Moreover, because it has the same form as a coherent condensate state, it also put the 
concept of vacuum as a state without particle and vacuum as the minimum of the potential of a 
condensate into the same formal formulation. This unified description is notably interesting in 
the context of symmetry breaking and phase transition. We leave a more extended study of 
$|\Psi_{G}\rangle$ and its properties to a future work.

\subsection{Vacuum symmetry} The issue of the vacuum energy can be considered yet in 
another way. When operator $a_k^\dagger$ is applied to the vacuum, one particle with momentum 
$\vec{k}$ is produced. But, because we exactly know its momentum, there is no information 
about its position. Therefore, an observer who wants to apply the annihilation operator must 
first somehow localize the particle, otherwise the probability of annihilation becomes 
negligibly small. Such an operation would not be possible without breaking the translation 
symmetry of the spacetime, for instance by imposing boundaries at a distance $L \sim 1/k$. 
This induces a Casimir energy proportional to $1/L \sim k$, thus infinite energy for 
$k \longrightarrow \infty$ i.e. if we want to be sure that the particles is 
annihilated.\footnote{This argument also shows that the Casimir effect~\cite{casimirobs,casimirobs0} 
produced by small anisotropies only add a finite contribution to the infinity obtained in 
(\ref{t00minkovski}). In fact, it adds a random component, because fluctuations are random. 
Therefore it cannot explain the presence of a uniform dark energy.} This description and the 
definition of vacuum as a coherent state given in the previous section are different view of the 
same reality, i.e. the close relation between quantum mechanics, its non-locality, and symmetries. 
The effect of particle creation given above explicitly shows the manifestation of this 
non-locality. Therefore, introduction of ordering operator or any other 
regularization-renormalization prescription can be considered as a shorthand for the operation 
described above i.e. creation of a particle with a given momentum in the Minkowski spacetime, 
limiting the space which needs/produces energy, annihilating the particle, and finally removing 
boundaries which produces/needs energy~\footnote{For the sake of simplicity here we only discussed 
the case of a Minkowski spacetime. Our arguments applies equally to other spacetimes, notably FLRW 
and de Sitter which are important in the cosmological context. If an argument is specific to one 
of these spacetimes, it is explicitly reminded in the text.}.

Some authors~\cite{weinbergde} have considered the infinity or very large vacuum energy 
obtained in (\ref{t00minkovski}) as a physical reality, specially because of the observation 
of Casimir effect~\cite{casimirobs,casimirobs0}. Thus, they have tried to find a way to neutralize 
vacuum energy by considering supersymmetry or quantum cosmology (see ~\cite{weinbergde} for review and 
references therein). Casimir effect is observable when the symmetry of an empty spacetime is 
explicitly broken, for instance by introducing a boundary which divides the space. As we have 
argued above, any symmetry breaking operation changes the quantum state of the spacetime, 
notably from a vacuum to a {\it non-vacuum}. One can define a vacuum as the state with smallest 
information - entropy. This is a direct conclusion of definition (\ref{condwaveg}). 
Every symmetry breaking induces some information. In the example above the division of space 
in two parts creates information because modes/particles are assumed to belong to one or the 
other part not both. Moreover, the confinement of particles in space is not a trivial operation 
and a {\it wall} or potential - thus energy - is needed to prevent particles to penetrate to the 
other part(s). This close relation between the way an experiment is performed and its outcome is a 
well known concept in quantum mechanics. Furthermore, inconsistencies of a semi-classical 
treatment of gravity are well known~\cite{grsemiqmincon,houriqgr}. A very explicit conflict 
between two theories, which is also directly related to the subject of vacuum energy, is the fact 
that by defining an absolute energy reference we lose all information about the position of 
the reference object.~\footnote{Note that this argument is not in conflict with 
$E_{min} = 0$ in supersymmetric models. The statement in these models is that the potential 
energy of all particles and their supersymmetric partner is zero. Although one can conclude 
that in SUSY models it is not possible to have a negative energy, in what concerns measuring 
the potential the laws of quantum mechanics are applied, i.e. if we want to verify $E_{min}$ 
theorem by measuring its value with very good precision, we lose information about the position of 
particles.}

Summarizing the conclusion of arguments given in this section, the regularization of energy-momentum 
tensor in quantum field theory is a necessary and legitimate technique for obtaining a physically 
meaningful value for this quantity. Accepting this point, it would be interesting to have a 
quantitative estimation of the energy density of vacuum after regularization for some physically 
important spacetimes. It has been shown~\cite{qftcurve} that renormalized energy momentum tensor 
$\langle T^{\mu\nu}\rangle_{ren}$ in de Sitter space is $\propto R_{dS}^{-4}$ where $R_{dS}$ is 
the radius of the de Sitter space. Assuming an initial radius $\sim H_{inf}^{-1}$ for an $\alpha-$vacuum, 
$H_{inf} \lesssim 10^9$ GeV, and $\gtrsim e^{60}$ folds expansion, the vacuum energy at the end of 
inflation must be $\lesssim 10^{-32}$ eV$^4$ which is much smaller than the observed energy density 
of dark energy $\rho_{de} \sim 10^{-11}$ eV$^4$.

In a similar way, one can formulate and apply regularization-renormalization to the expectation 
value of energy density in other - less symmetric - curved spacetimes such as FLRW. Dynamic 
equation in FLRW do not always have an analytical solution, but approximate solutions have been 
found in~\cite{houricondense}. In particular, in flat-FLRW geometry equal-time space-like surfaces 
have the same geometry as their counterpart in Minkowski space. Thus, crossing singularity and its 
regularization has the same properties too. The dominance of dark energy and accelerating 
expansion of the Universe at present makes the spacetime to approach to a de Sitter geometry. 
Thus, using the same relation between $\langle T^{\mu\nu}\rangle_{ren}$ and horizon size 
$\sim H_0^{-1}$ should give an acceptable order of magnitude estimation for an accelerating FLRW 
geometry\footnote{A heuristic way of reasoning is to say that in a bounded 4-dim empty spacetime 
of size $R$, according to Heisenberg uncertainty principle, the uncertainty on the energy density 
of quantum fluctuations is $\sim R^{-4}$.}. This estimation leads to a very small vacuum energy 
of $\lesssim 10^{-120}$.

In the rest of this letter we discuss some of other arguments in the literature about the physical 
reality of very large vacuum energy according to quantum field theory.

\section{Gravitational coupling of virtual particles} \label{sec:virtual}
The running mass of elementary particles such as electrons and observed phenomena such as Lamb shift 
which are due to quantum fluctuations, in another word interaction with virtual particles from 
vacuum, have been considered as an evidence in favour of a large nonzero vacuum energy obtained in 
equation (\ref{t00minkovski}), see~\cite{weinbergew} and references therein, 
and~\cite{devacuumrev,devacuumrev0}. The claim is that the modification of mass and coupling 
constants due to loop corrections i.e. exchange of virtual particles, is the evidence of coupling 
between the vacuum and gravitons, and thereby gravitational interaction of vacuum. Therefore, the 
observed small nonzero density of dark energy is concluded to be an evidence for the failure of 
quantum field theory in what concerns its prediction for gravitational interactions. 
Both scale dependence of mass and Lamb shift are well understood and measured phenomena. Giving the 
fact that gravitational interaction is proportional to mass/energy, it is evident that the 
interaction of gravitons with virtual particles from vacuum is included in the mass of elementary 
particles. To obtain the relation between energy scale and effective mass e.g. in the context of 
the Quantum Electro-Dynamics (QED) one has to renormalize the theory by removing singularities 
similar to what obtained in (\ref{t00minkovski}). The fact that after this apparently ad hoc 
operation we obtain relations that are confirmed by experiments, proves that these ad hoc 
calculations are after all meaningful. If the argument above about a large vacuum energy was true, 
for the same high energy cutoff, diagrams with larger number of loops should have a larger effect 
on the effective mass and coupling of electrons at scales less than the cutoff. Thus, they should 
make a model like QED completely unpredictive. Indeed, the scale dependence of mass and couplings 
demonstrates that in the context of QED, the effect of virtual particles is controlled by the 
electromagnetic coupling rather than by the energy-momentum, because with increasing number of 
loops the small coupling of QED decreases their effect despite the fact that every loop includes a 
singular momentum integral.

In fact, this effect can be explicitly demonstrated if in analogy with gravity we consider 
semi-classical electrodynamics in which charged particles are considered to be quantum fields but 
the electromagnetic field is treated classically. This example is interesting because we know very 
well both the classical and fully quantic models for electromagnetic interaction in Minkowski 
spacetime. The semi-classical Maxwell equation can be written as:
\be
F^{\mu\nu}_{,\nu}(x) = \langle 0|j^\mu(x)|0 \rangle, \quad \quad j^\mu \equiv \bar{\psi} 
\gamma^\mu \psi \label{emsemi}
\ee
To determine the right hand side of (\ref{emsemi}) at zero-order, we can decompose the fermion 
field $\psi$ to modes similar to (\ref{commutrel}) with only difference that they anticommute. It 
can be easily shown that this leads to an integral similar to what is in (\ref{t00minkovski}) with 
a linear singularity due to nonzero anticommutation of modes with the same momentum. Similar to 
determination of $T^{\mu\nu}$, if ordering operator is applied the singularity will be removed. 
There are also higher order corrections to the vacuum current. If they produce a nonzero current 
in the right hand side of (\ref{emsemi}), the classical electromagnetic field would be also nonzero 
and interacts as an external field with virtual particles. The gravitational analogue of such a 
back-reaction has been claimed to be the reason for a large vacuum energy density - see diagrams 
in figure 2 of~\cite{devacuumrev} for examples of interaction between graviton and virtual 
particles. As explained in the previous paragraph, it is claimed that they generate a very large 
vacuum energy. Fortunately, in the case of electrodynamics we have a full renormalizable quantum 
field theory and we know that after regularization and renormalization, the right hand side of 
equation (\ref{emsemi}) becomes null, thus QED vacuum has no charge or current. It can be claimed 
that the neutrality of QED vacuum is due to charge conjugate symmetry. In the neutrino sector of 
electroweak CP is violated. However, no difference between propagation of neutrinos and 
anti-neutrinos in vacuum has been observed~\cite{nupropagvacuum}. Recent claims about observation 
of a slight difference between neutrinos and anti-neutrinos appearance in short baseline oscillation 
experiments~\cite{nulsnd,numiniboone,numiniboone0} is most likely related to mixing with non-Standard 
Model sterile neutrinos, see e.g.~\cite{nunubardiff} or uncertainties of anti-neutrinos initial 
flux~\cite{numiniboonflux,numiniboonflux0}, see also below for issues related to electroweak vacuum. 

We should remind that the argument in~\cite{devacuumrev} about the interaction of graviton 
with virtual particles is misleading. In Standard Model mass is generated as a result of 
interaction (see also the next section for issues concerning the condensation of Higgs). Therefore, 
it can be considered as the effect of interactions - a delay - in the propagation of a particle. 
As a classical theory, virtual particles {\bf do not exist} for Einstein gravity. Only in the 
context of a quantum gravity - which is not available - an interaction between graviton and 
virtual particles can be envisaged. In Sec. \ref {sec:newinterp} we defined the vacuum as a 
coherent state with vanishing amplitude for all momentum modes. Considering the argument above 
and the fact that quantum effects of vacuum can be only detected in presence of quantum 
interactions, we can say that even in the context of quantum field theory vacuum is the empty 
space without any external effect and what we call vacuum excitations are in fact spontaneous 
excitation of fields/particles. This interpretation is more consistent with procedure of 
renormalization and does not leave any ambiguity such as whether graviton interact with vacuum or 
not. 

Another argument in support of the validity of renormalization in curved spacetimes is the fact 
that predictions of quantum field theory and SM physics are observed and tested in strong gravity 
field regime around compact objects such as neutron stars, pulsars, and accretion disk of black 
holes, see e.g.~\cite{nstarbhpred}. If the apparent infinities of energy-momentum tensor and 
other physical quantities before their renormalization were physical, we could not make any 
prediction for the observed quantities such as modification of emission lines and radiation transfer 
processes in the accretion disk around black holes in AGNs~\cite{bhlines} and compact binary 
stars~\cite{bstaracc}, the rate of energy loss and slow-down of pulsars by gravitational wave 
production observed through their radio emission which is produced by electromagnetic 
interactions in a strong gravity field environment~\cite{pulsarem}, etc.

\section{Contribution of vacuum expectation value of the Standard Model 
condensates in dark energy} \label{sec:higgs}
One of the arguments usually raised in the literature in favor of a large vacuum energy density 
is the nonzero vacuum expectation value (vev) of Higgs boson and composite fields 
responsible for chiral symmetry breaking in QCD. The latter case has been recently discussed 
in~\cite{qcdchiralvacu}, and it is argued that the confinement of quarks in hadrons also confines 
their condensate $\langle \bar{q}q\rangle$ to the volume occupied by them rather than being 
homogeneously distributed in space. Therefore, we do not discuss this here. Nonetheless, our 
arguments in this section about the contribution of known condensates of elementary particles in 
dark energy are general and apply to quark condensates as well.

The most important condensate expected in the framework of the Standard Model is Higgs with a 
nonzero vacuum expectation value expected to be $\sim 246$ GeV. It is necessary 
for generating masses of Standard Model particles, in particular the mass of $W^\pm$ and $Z_0$ gauge 
bosons. In addition, it induces the breaking of $SU(2) \times U(1)$ symmetry at an scale 
$\gtrsim 1 TeV$. One should remind that here the word {\it vacuum} does not mean {\it empty} space 
but the minimum of the effective potential of Higgs field(s). The vev can be considered as a 
classical field, and in analogy with condense matter it is called a condensate. Like other 
quantities in quantum field theory, it is also subjected to regularization and renormalization. 
The effective self-interaction potential of the Standard Model Higgs is supposed to have two 
nonzero minima. The condensation of one of the components of Higgs doublet at one of these minima 
breaks $SU(2) \times U(1)$ gauge symmetry and generates mass for $W^\pm$ and $Z_0$ bosons. It is 
usually assumed that the Higgs condensate has a uniform distribution in space. Then, the question arises 
why the energy density of dark energy - which as a homogeneous density must include the vacuum 
energy density of Higgs and other scalars such as modulies in string theory - is $\lesssim 10^{56}$ 
folds smaller than Higgs vev~\cite{devacuumrev,devacuumrev0}. 

Here we argue that this apparent inconsistency is due to the confusion that the energy density 
of a condensate must be uniform. Conditions for the formation of a Bose-Einstein Condensate (BEC) 
in a fluid is studied in~\cite{beccond}. The process of condensation of a scalar field in FLRW 
spacetimes has been discussed in details in~\cite{houricondense}. A condensate is defined by 
decomposing the scalar field:
\be
\Phi = \phi + \varphi, \quad \quad \langle \phi \rangle = 0, \quad \quad v \equiv \langle \varphi 
\rangle \neq 0 \label{conddef}
\ee
It is the component with nonzero expectation value that behaves similar to a classical scalar 
field - a condensate. In both classical fluid and quantum field theory description of a condensate  
the amplitude of anisotropies decreases very rapidly for large modes if there is no addition 
driver to generate large mode fluctuation such as an external force in a fluid, see 
e.g~\cite{becextforce} and references therein, or in the case of quantum scalar fields an 
interaction with other fields~\cite{houricondense}. This 
observation is analogue to infinite volume condition for symmetry breaking in statistical physics. 
Note that in contrast to the case of symmetry breaking in which multiple fields and an internal 
symmetry are assumed, the results in~\cite{houricondense} are obtained for single field produced 
by the decay of another field without any assumption or constraint on their symmetry. The setup 
in which the scalar has other interactions applies to Higgs condensate. In general, renormalization 
of the underlying theory induces scale dependence to masses - including Higgs mass - and to 
couplings. Electroweak interaction is not asymptotically free, therefore couplings increases with 
energy scale. This means that particles, their interactions, and thereby their condensate are 
concentrated to short distances.

Another fact to be considered is the feedback of the formation of a condensate on its own evolution 
and on the evolution of other fields that interact with it. In particular, Higgs condensate 
interacts with gauge bosons and quarks, and generates mass for them, at least for gauge bosons, 
thus reduces their dynamism. Moreover, formation of a condensate does not necessarily break 
symmetries because it can simply uplift the potential. Only when the effective mass becomes 
negative the $Z_2$ symmetry of the Higgs potential will break. Therefore, if Higgs is responsible 
for the breaking of $SU(2) \times U(1)$ symmetry, it must have significant interactions with other 
fields, specially fields related to high energy physics - short distance scales - such that the 
negative mass condition be satisfied. Because the process of condensation is very closely related to 
interactions, one expects that the condensate distribution has strong correlation with these 
fields. For Higgs, fields to which it is coupled are: quarks that are confined to atomic distances, 
high energy physics constituents which are by definition confined to short distances, and gauge 
bosons $Z_0$ and $W^\pm$ which interact with it through $\propto H^2 A_\mu A^\mu$ term in SM Lagrangian. 
They have a short free path due to their short lifetime. Detailed study of a simpler case, the 
condensation of a scalar field produced by the decay of a heavy particle after inflation and during 
reheating in~\cite{houricondense}, shows that the condensate can be easily confined by interactions 
if their couplings to heavy particles are dominant at short distance scales, see equations 
(84-85) in~\cite{houricondense}. Some of Higgs models consider strong interactions for Higgs, 
either QCD interaction or non-SM interactions with strong couplings~\cite{higgsstrongint}. Even if 
Higgs has only electroweak interaction, according to these arguments the increase of electroweak 
interaction coupling with scale~\cite{weinbergew} alone can be sufficient to confine the condensate 
to short distances. The absence of a large uniform vacuum energy density due to condensation of 
Higgs or other fundamental scalar fields such as scalar fields in SUSY models or modulies in 
string theory models - expected to live in higher energy scales - means that their condensates are 
confined to short distances. Therefore, their effects are included in the renormalized mass and 
couplings of particles observed at low energies. 

We should remind that confinement is also present in superconductivity and superfluidity - the only 
places where a Higgs phenomenon is actually observed. The analogy between condensation of 
fundamental fields and what is observed in condense matter is also discussed 
in~\cite{qcdchiralvacu}. In condense matter the complex scalar is a composite Cooper pair. The 
interaction that forces electrons to make a spin-zero pair is the effective electromagnetic force 
of the lattice of ions presented by its pseudo-particle vibrational modes called phonons. The 
self-interaction of Cooper pairs is usually considered to be $\phi^4$ type with a global $Z_2$ 
symmetry which is broken by the condensate. The latter is confined to where Cooper pairs are 
present. In this analogy the lattice plays the same role as high energy physics plays for Higgs. 
Although one of the necessary conditions for formation of a condensate, the symmetry breaking, and 
the phase transition is a continuous scalar field in an infinite volume, we know that Cooper pairs 
and their condensate are limited to the volume of a superconductor in the lab. Thus, if the 
analogy with Higgs condensate is correct, and if the mass and self-interaction of the SM Higgs is 
related to physics at high energy scales, we expect that its nonzero expectation value is confined 
to very short distances. Therefore, at low energies its effect is observable only through 
renormalized properties of particles. In the case of quintessence models for dark energy, the 
survival of the quintessence field condensate at cosmological scales is a consequence of its very 
small mass and very weak coupling that leads to the formation of a coherent state which is close to 
uniform at cosmological distances and survives the expansion of the Universe~\cite{houricondense}.

Our claim about the confinement of Higgs condensate at high energies can be tested by measuring 
the energy density of the vacuum at scales close to its expectation value. For instance, considering 
the interaction between a Higgs doublet and $SU(2) \times U(1)$ and symmetry breaking by the Standard 
Model Higgs condensates, Feynman diagrams can include zero, one, or two contribution from the 
condensate. The latter case generates mass for gauge fields. The lowest order effective interaction 
between massive gauge bosons and the quantum Higgs field is 
$\propto (m^2_A / v) H A^\mu A_\mu$, see e.g. ~\cite{higgsprop}. Variation of the cross-section of these 
events with energy scale measures energy dependence or equivalently the spatial extend of the Higgs 
condensate in space, similar to observation of confinement in strong interaction and scale dependence 
of QCD coupling. If the signature of Higgs claimed by the LHC 
Collaboration~\cite{higgslhc,higgslhc0,higgslhc1} is confirmed, accumulation of data in near future 
will make it possible to measure the spatial distribution of the Higgs condensate.

\section{Outline}
We conclude that ordering operation or another regularization/renormalization method are necessary 
operations to 
make the definition of energy-momentum tensor consistent with quantum mechanics and quantum field 
theory principles. They are not ad hoc prescriptions to make the vacuum energy finite when the zero 
point of energy can be chosen arbitrarily. The apparent ambiguity in the definition of energy 
is closely related to quantum uncertainties, nonlocal nature of quantum mechanics, and its relation 
with symmetries. The latter topic needs clarifications that we leave to another work. 

We presented a new definition for vacuum as a complete coherent state that includes all states in the 
Fock space of the Universe with a negligible probability. An observer can only detect a state when 
it is decohered from this maximally coherent state. It unifies the definition of vacuum as the minimum 
of the potential and vacuum as a particle-less state. Such a state can play the role of a 
{\it quantum background}. Although it satisfies the usual definition of vacuum state with zero 
expectation value for all fields, it shows explicitly that quantum vacuum is not as empty as a 
classical vacuum is. One of its notable properties is its frame-independence. Therefore, in this 
respect it is unique. It also provides a quantum mechanical discrimination between {\it real} and 
{\it virtual} particles without referring to frame and observer dependent quantities such as energy 
and momentum. It would be interesting to see whether this definition of vacuum can play the role 
of a background in some of quantum gravity models in which there is no apparent background. 

Using this vacuum state and analogy between semi-classical gravity and semi-classical 
electrodynamics, we showed that the apparent large or infinite energy density of vacuum is not a 
physical observable and can be safely regularized. We discussed the issue of Higgs large vacuum 
expectation and argued that it is confined to scales of the order of electroweak symmetry breaking 
scale. These explanations persuade us to reconsider the interpretation of dark energy as the 
{\it energy of vacuum}, and encourage explanations based on either condensation of fields - 
quintessence models - or a modification of gravity. Notably, in some classes of quintessence models 
the condensate density can be very uniform and has an equation of state very close to a 
cosmological constant~\cite{houridmquin}. Only observations of additional evidence such as 
interacting/decaying dark matter, and interaction in the dark sector can discriminate these models 
from a genuine cosmological constant~\cite{houridiscrim}.


\begin{thebibliography}{10}
\bibitem {descalar} E.J. Copland, M. Sami, \& S. Tsujikama, \Journal {\IMD}{15}{2006}{1753} [hep-th/0603057].
\bibitem {devacuumrev} J. Polchinski [hep-th/0603249].
\bibitem {devacuumrev0} R. Bousso [arXiv:1203.0307].
\bibitem {qcdchiral} A.A. Belavin, \etal, \Journal{\PLB}{59}{1975}{85}.
\bibitem {qcdchiral0} G. t'Hooft, \Journal{\PRD}{14}{1976}{3432}.
\bibitem {qcdchiral1} A. Casher \& L. Susskind, \Journal {\PRD}{9}{1974}{436}.
\bibitem {qcdchiralvacu} S.J. Brodsky \& R. Shrock, \Journal{\PNS}{}{2011}{45}.
\bibitem {adst00} M. M. Caldarelli, \Journal{\NPB}{549}{1999}{499}, [hep-th/9809144].
\bibitem {weinbergde} S. Weinberg, \Journal {\RMP}{91}{1989}{1}.
\bibitem {qftcurve} N.D. Birrell \& P.C.W. Davis, {\it Quanyum fields in curved space}, Cambridge Univ. Press, (1986).
\bibitem {stringlandscape} E. Witten, [hep-ph/0002297].
\bibitem {stringlandscape0} J. Kumar, \Journal{IMA}{21}{2006}{3441} [hep-th/0601053] (review).
\bibitem {antropologic} S. Weinberg, \Journal{PRL}{59}{1987}{2607}.
\bibitem {antropologic0} A. Vilenkin, \Journal{\PRL}{74}{1995}{846}.
\bibitem {semiqmgrincon} K. Eppley, \& E. Hannah E., \Journal{\FOP}{7}{1977}{51}.
\bibitem {houriqgr} H. Ziaeepour, \Journal{\JPC}{174}{2009}{012027} [arXiv:0901.4634].
\bibitem {condwave} S. Matsumoto and T. Moroi \Journal {\PRD}{77}{2008}{045014}, arXiv:0709.4338.
\bibitem {houricondense} H. Ziaeepour, \Journal {\PRD}{81}{2010}{103526} [arXiv:1003.2996]
\bibitem {exciton} M. Kira, G. Khitrova, \& H.M. Gibbs, \Journal{\NAM}{5}{2006}{523}.
\bibitem {casimir} H.B.G. Casimir, {\em Proc. K. Ned. Akad. Wet}, {\bf 51} (1948) 793.
\bibitem {casimirobs} S.K. Lamoreaux, \Journal{\PRL}{78}{1997}{5}.
\bibitem {casimirobs0} G.L. Klimchitskaya, \etal, \Journal{\IMA}{}{2005}{1}.
\bibitem {pulsar} Taylor, J.H., Fowler, L.A., \& Weisberg, J.M., \Journal{\NAT}{277}{1979}{437}.
\bibitem {nstarbhpred} S. Vaughan, A. C. Fabian, \Journal {\MRA}{341}{2003}{496} [astro-ph/0301172].
\bibitem {desittervac} B. Allen, \Journal {\PRD}{32}{1985}{3136}.
\bibitem {bunchdavis} T.S. Bunch \& P.C.W. Davis, \Journal{\PRS}{360}{1978}{117}.
\bibitem {alphavac} M.B. Einhorn \& F. Larson [hep-ph/0305056].
\bibitem {grsemiqmincon} Wald, R.M., The University of Chicago Press, (1984).
\bibitem {weinbergew} S. Weinberg, {\it The Quantum Theory of Fields} vol. 2, Cambridge University Press (1997). 
\bibitem {nupropagvacuum} A.A. Aguilar-Arevalo, \etal, [arXiv:1109.3480].
\bibitem {nulsnd} A.A. Aguilar-Arevalo, \etal, \Journal{\PRD}{64}{2001}{112007} [hep-ex/0104049].
\bibitem {numiniboone} A.A. Aguilar-Arevalo, \etal, \Journal{\PRL}{98}{2007}{231801} [arXiv:0704.1500].
\bibitem {numiniboone0} A.A. Aguilar-Arevalo, \etal, \Journal {\PRL}{105}{2010}{181801} [arXiv:1007.1150].
\bibitem {nunubardiff} E. Akhmedov, T. Schwetz, \Journal {\JHE}{1010}{2010}{115} [arXiv:1007.4171]. 
\bibitem {numiniboonflux} A.A. Aguilar-Arevalo, \etal, \Journal{\PRD}{84}{2011}{072005} [arXiv:1102.1964].
\bibitem {numiniboonflux0} G. Mention, M. Fechner, Th. Lasserre, Th. A. Mueller, D. Lhuillier, M. Cribier, A. Letourneau, \Journal{\PRD}{83}{2011}{073006} [arXiv:1101.2755]. 
\bibitem {bhlines} J. Wilms, \etal, \Journal {\MRA}{328}{2001}{L27} [astro-ph/0110520].
\bibitem {bstaracc} L. Angelini, N.E. White, \Journal {\APJ}{586}{2003}{L71} [astro-ph/0302315].
\bibitem {pulsarem} S. Zane, \etal, \Journal {\MRA}{334}{2002}{345} [astro-ph/0203105].
\bibitem {beccond} O. Penrose \& L. Onsager, \Journal{\PRV}{104}{1956}{576}.
\bibitem {becextforce} M.L. Chiofalo \& M.P. Tosi, \Journal {\EPL}{56}{2001}{326}.
\bibitem {higgsstrongint} G.F. Giudice, C. Grojean, A. Pomarol, R. Rattazzi, \Journal {\JHE}{0706}{2007}{045} [hep-ph/0703164].
\bibitem {higgsprop} I. Low, J. Lykken, \Journal{\JHE}{10}{2010}{053} [arXiv:1005.0872].
\bibitem {higgslhc} ATLAS Collaboration, \Journal {\PLB}{710}{2012}{49}.
\bibitem {higgslhc0} ATLAS Collaboration, \Journal {\PRL}{108}{2012}{111803} [arXiv:1202.1414].
\bibitem {higgslhc1} ATLAS Collaboration, \Journal{\PLB}{710}{2012}{383} [arXiv:1202.1415].
\bibitem {houridmquin} H. Ziaeepour, \Journal {\PRD}{69}{2004}{063512} [astro-ph/0308515].
\bibitem {houridiscrim} H. Ziaeepour, \Journal{\MPL}{22}{2007}{1569} [astro-ph/0702519].
\bibitem {houridiscrim} H. Ziaeepour, \Journal{\AIP}{957}{2007}{453} [arXiv:0709.0115], 
\bibitem {houridiscrim} H. Ziaeepour, to be published in \PRD, 30 July (2012) [arXiv:1112.6025].
\end{thebibliography}
\end{document}